# Mirror triad of tectonic earthquakes

O.D. Zotov, A.V. Guglielmi

*Schmidt Institute of Physics of the Earth, Russian Academy of Sciences, Moscow, Russia,*
*ozotov@inbox.ru , guglielmi@mail.ru*

**Abstract**

It is natural to call the trinity of foreshocks, main shock and aftershocks a classical triad. The magnitude of the main shock is always greater than the maximum magnitudes of foreshocks and aftershocks. In the classical triad, the number of foreshocks is less than the number of aftershocks. In many cases, foreshocks are completely absent even before very strong earthquakes. In this work, we found a rare but extremely interesting species of triad (we called it the mirror triad), in which the number of foreshocks is greater than the number of aftershocks at the interval of 24 h before and 24 h after the main shock. Quite often, aftershocks are absent in mirror triads at all. In this paper, we described the properties of mirror triads and made an attempt to understand their origin. A schematic classification of tectonic triads is proposed.

### 1. Introduction

Apparently, the idea of a peculiar trinity of foreshocks, main shock and aftershocks in a sequence of tectonic earthquakes [1–3] was formed in seismology not without the influence of mathematics, in which a binary relation between elements of a set can give rise a trichotomic relation. The trinity of foreshocks, main shock and aftershocks was proposed to be called the classical triad [4]. The magnitude of the main shock $M_0$ is always greater than the maximum magnitudes of foreshocks and aftershocks. The classical triad satisfies the inequalities

$$M_- < M_+, \qquad (1)$$

and

$$N_- < N_+. \qquad (2)$$

Here $M_-$ ($M_+$) and $N_-$ ($N_+$) are the maximum magnitude and the number of foreshocks (aftershocks), respectively.

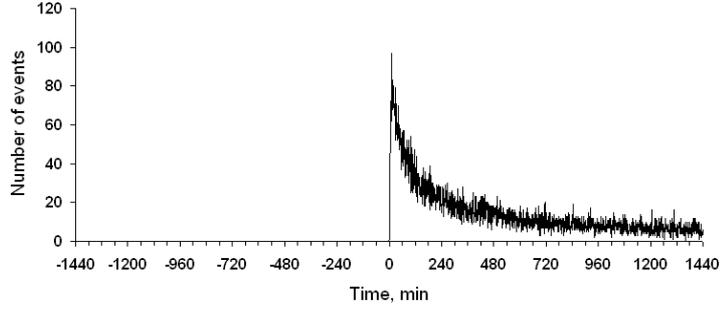

Fig. 1. Generalized picture of a shortened triads. Zero moment of time corresponds to the moment of the main shock.

Quite often $N_- = 0$, i.e. foreshocks are absent even before rather strong earthquakes. Figure 1 illustrates this situation (the database and the plotting method will be described in detail in the next section of the paper). With regard to aftershocks, there is a stable opinion that after a sufficiently strong earthquake, repeated tremors are always observed, i.e. $N_+ \neq 0$.

The purpose of this paper is to present a rare but extremely interesting type of triad (we call it the mirror triad), for which inequalities that are directly opposite to inequalities (1) and (2) hold. Moreover, $N_+ = 0$ in a significant part of the mirror triads.

## 2. Mirror triad

An extensive literature is devoted to the experimental study of the classical triads. We point here to work [5], since in the study of mirror triads we used a database and general methods of analysis similar to those used in [5] in the study of classical triads (see also [6]).

We used data on earthquakes that occurred on Earth from 1973 to 2019 and were registered in the world catalog of earthquakes USGS/NEIC (https://earthquake.usgs.gov). There were found 2508 main shocks with a magnitude of $M_0 \geq 6$ and a hypocenter depth not exceeding 250 km. For each main shock, a circular epicentral zone was determined by the formula

$$\lg L = 0.43 M_0 - 1.27, \qquad (3)$$

where the radius of the zone $L$ is expressed in kilometers [7]. According to our definition, the classical triad is formed by earthquakes, which occurred in the epicentral zone in the interval of $\pm 24$ hours relative to the moment of the main shock provided that the inequalities (1), (2) are satisfied. The total number of earthquakes was distributed among the members of the triad in the following way: $N_- = 1105$, $N_0 = 2398$, $N_+ = 31865$. Figure 1 is based on truncated triads, in which there are no foreshocks. The graph was constructed by the method of overlapping epochs, and the main shock of the earthquake was used as a benchmark. For truncated triads, the distribution looks like this: $N_- = 0$, $N_0 = 2066$, $N_+ = 21422$. The distribution at $N_- \neq 0$: $N_- = 1105$, $N_0 = 332$, $N_+ = 10443$. We see that in the presence of foreshocks, the activity of aftershocks is higher than in the absence of foreshocks.

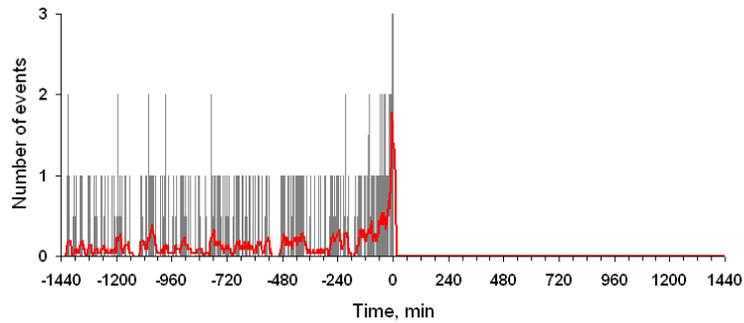

Fig. 2. Generalized view of truncated mirror triads. Zero moment of time corresponds to the moment of the main shock. The red line is obtained by averaging by moving average (window 20 min, step 1 min).

In the course of the study of classical triads, one of the authors (O.D.) put forward the idea of making a selection of observation data, replacing inequalities (1), (2) with the opposite ones. As a result, it was possible to find the mirror triads. Figure 2 shows the truncated mirror triads: $N_- = 237$, $N_0 = 156$, $N_+ = 0$. If $N_+ \neq 0$, then $N_- = 1375$, $N_0 = 104$, $N_+ = 755$.

We see that the mirror triad are relatively rare phenomenon. They appear about an order of magnitude less frequently than classical triads. To make the picture of mirror triads more visual, we will show Figure 3. It shows mirror triads in the range of main shock magnitudes $M_0 = 5-6$. Here $N_- = 4189$, $N_0 = 2430$, $N_+ = 201$.

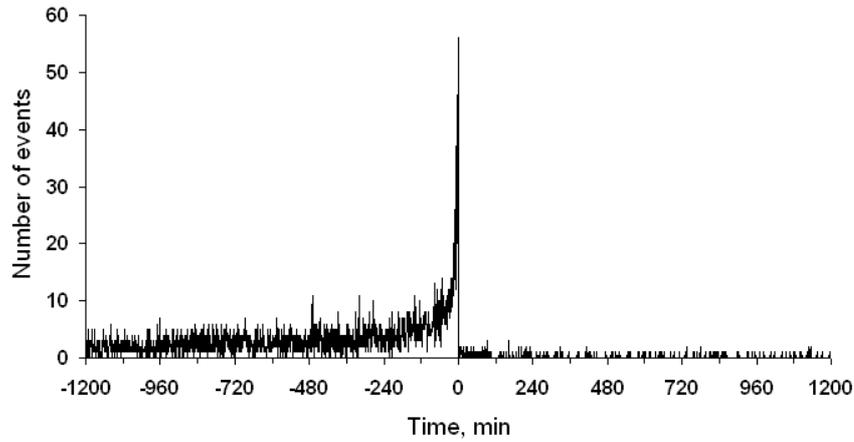

Fig. 3. Time distribution of foreshocks and aftershocks of mirror triads in the range of magnitudes of the main shocks $M_0 = 5-6$.

So, we found that there is a rare but rather interesting subclass of tectonic earthquakes with magnitudes $M_0 \geq 6$, in which the number of aftershocks in the interval of 24 h after the main shock is significantly less than the number of foreshocks in the same interval before the main shock. In many cases, there are no aftershocks at all. We asked the question: Are there earthquakes with magnitudes $M_0 \geq 6$, neither before nor after which there are neither foreshocks nor aftershocks? The search result was amazing. We have discovered a wide variety of this kind of earthquake and named it *Grande terremoto solitario*, or GTS for short. In Figure 4, we see that the number of GTS (2460) is approximately equal to the number of classical triads (2398).

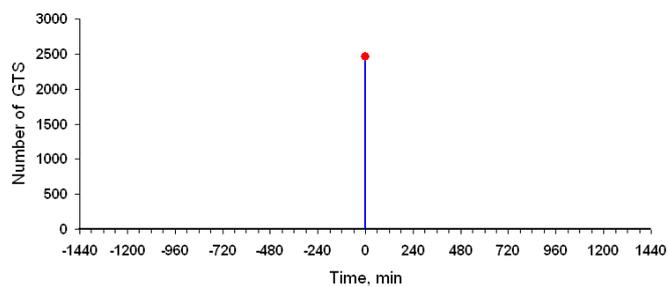

Fig. 4. Solitary earthquakes with magnitudes $M_0 \geq 6$.

GTS arise spontaneously under very calm seismic conditions, and are not accompanied by aftershocks. This suggests an analogy between the GTS and the so-called "Rogue waves" (or f "Freak waves") – isolated giant waves that occasionally emerge on a relatively quiet ocean surface (e.g., see [8]). This analogy may prove to be quite profound, since the spontaneous occurrence of pulses having anomalously high amplitudes is a common property of the nonlinear evolution of dynamic systems [9].

For completeness, we also present the data for the case $N_+ = N_-$: $N_- = 186$, $N_0 = 121$, $N_+ = 186$. The triad in which $N_+ = N_-$ will be called symmetric. It is interesting to note that formally GTS can be related to a variety of symmetric triads, since for them $N_- = N_+ = 0$.

As a summary, we point out that classical (normal) triads make up approximately 85% of all triads. Anomalous triads account for 15%, with mirrored 10% and symmetrical 5%. In this calculation, we excluded the GTS, which seem to form a special set of earthquakes.

## 3. Discussion

The detection of mirror triads is of interest for the physics of earthquakes. In this preliminary report, we only briefly discuss their properties.

The Gutenberg-Richter law [10] seems to hold true for foreshocks and main shocks of truncated triads. The law definitely executed for the main shocks of the mirror triads. In this paper, we restrict ourselves to giving the Gutenberg-Richter distribution for GTS: $\lg N = 10.6 - 1.4M$ (see Figure 5).

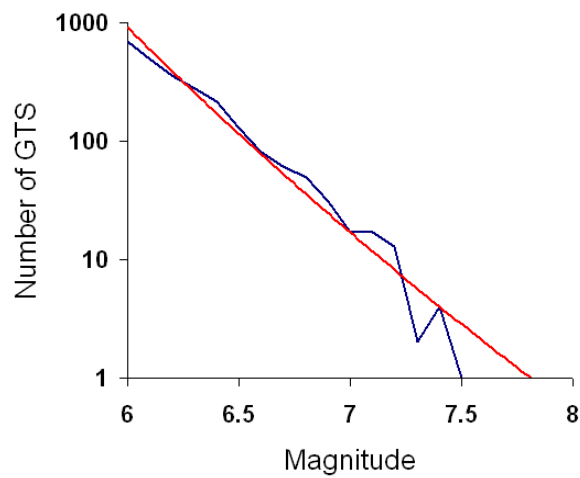

Fig. 5. Dependence of the number of GTS on the magnitude. The Gutenberg-Richter distribution is shown with a red line.

We noticed that some analogue of Bath's law also holds [11]. The difference $\Delta M$ between the magnitude of the main strike and the maximum magnitude of the foreshocks is no less than the value approximately equal to $\Delta M = 1$. However, the picture was fuzzy due to a relatively small number of events at $M_0 \geq 6$. A more definite result turned out for a set of events shown in Figure 3 ($5 \leq M_0 < 6$). The result is shown in Figure 6. We see that $\Delta M$ increases from 0.5 to 1 with increasing $M_0$ in the specified range. It should be emphasized that the question of an analogue of the Bath law for mirror triad needs further study.

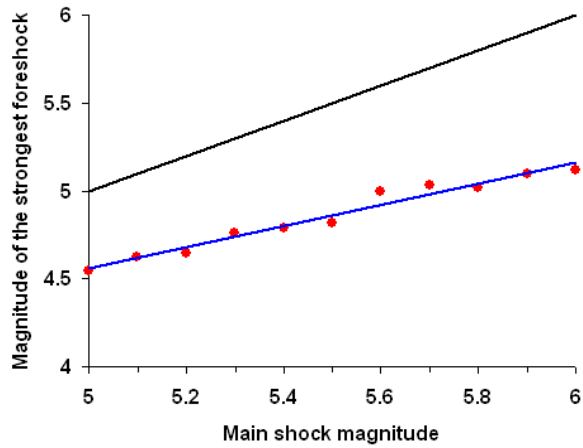

Fig. 6. Comparison of the magnitudes of the strongest foreshocks (points) and the magnitudes of the main shocks. The blue line approximates the experimental points. The black line is drawn for clarity.

Figures 2 and 3 show that foreshocks in the mirror triad appear to have a temporal distribution similar to the Omori distribution for aftershocks in the classical triad. Let's dwell on this in more detail. We represent the classical Omori law [12] in the simplest differential form

$$\dot{g} = \sigma_+. \qquad (4)$$

Here $g = 1/n$, $n(t)$ is the frequency of aftershocks, $t > 0$, the dot above the symbol means time differentiation, $\sigma_+$ is the so-called deactivation factor of the earthquake source, "cooling down" after the main shock (см. [5, 13, 14]). Suppose that for foreshocks of the mirror triad, the evolution law (4) is fulfilled with the replacement of $\sigma_+$ by $\sigma_-$. It is natural to call the $\sigma_-$ value the activation factor.

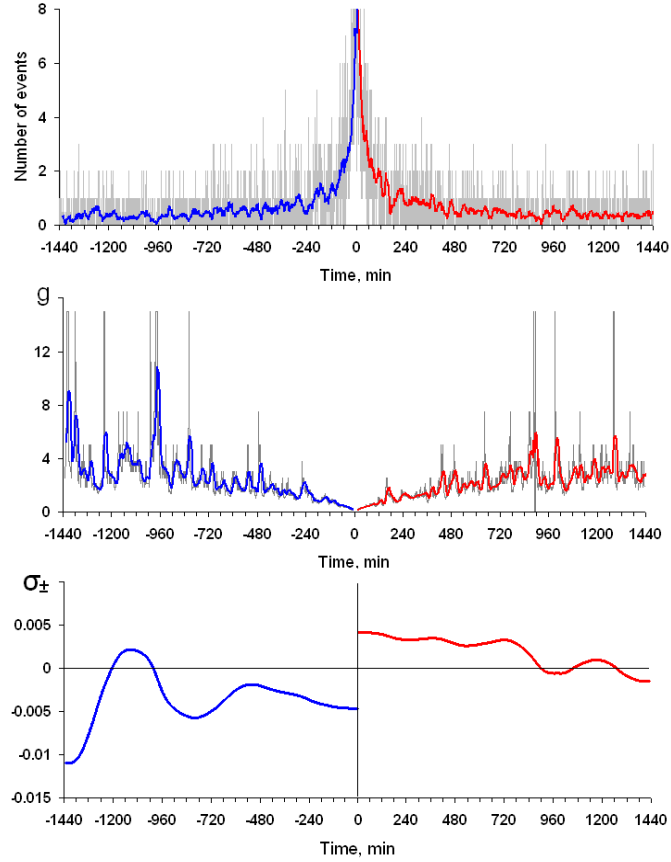

Fig. 7. Time dependence of the properties of symmetric triads. From top to bottom: earthquake frequency, auxiliary function, activation and deactivation factors.

In this paper, we will limit ourselves to presenting the interesting figure 7. It shows the generalized evolution of foreshocks and aftershocks in symmetric triads satisfying the condition $5 \leq M_0 < 6$. Here $N_- = 1050$, $N_0 = 742$, $N_+ = 1050$. The top two panels show an amazing mirror image. In the bottom panel, we have shown the variations of the $\sigma_-$ and $\sigma_+$ functions as the first step towards studying the activation and deactivation coefficients of an earthquake source in mirror triads. (For the procedure for calculating $\sigma_\pm$, see [14].)

In conclusion of this section of the paper, we would like, with all the necessary reservations, to express a careful judgment on the question of the origin of mirror triads. Let us assume that a system of faults in a certain volume of rocks is under the influence of a slowly growing total shear stress $\tau$. Threshold tension $\tau_*$ at which destruction occurs, i.e. the sides of the fault shift and a rupture occurs, generally speaking, the lower, the larger the linear dimensions of the fault $l$:

$$\tau_* = Cl^{-m}. \qquad (5)$$

Here $C$ is the dimensional coefficient of proportionality, depending on the properties of rocks in the selected volume, $m > 0$. Then the largest fault reaches the threshold first. Its destruction is manifested in the form of the main shock of an earthquake with a magnitude $M_0$.

If the $C$ parameter is uniformly distributed over the source volume, then foreshocks do not arise. Aftershocks appear due to the fact that after the main shock, the general external stress is partially removed, and the remaining stress is redistributed in a complex way throughout the volume. The local over-tensions arise, and in such a way that smaller faults than the one that generated the main blow can be activated and give repeated tremors. This is how we can imagine the emergence of a shortened classical triad.

In some cases, the specific distribution of faults in terms of $l$ and the distribution of local stresses may turn out to be such that not a single aftershock occurs. It is possible that such a situation occurs when the GTS is excited.

The appearance of the mirror triad can be understood if we assume that the parameter $C$ is not uniformly distributed over the volume, or rather, that there is a strong scatter in the $C(l)$ values. Then a situation is possible when, before the largest fault is destroyed, smaller faults are activated and foreshocks appear. A triad of tectonic earthquakes will appear. Whether it will be classical or mirror-like depends on the distribution of faults by the value of $l$, on the dispersion of the $C(l)$ coefficient, and on the mosaic of local stresses that arose after the main shock.

The above indicative considerations can be represented schematically in the form of a table:

Table. Classification of triads

| Local stresses | Scatter of $C(l)$ | Triad type |
|---|---|---|
| Small | Small | GTS |
| Large | Small | Classic truncated |
| Small | Large | Mirror truncated |
| Large | Large | Complete triad * |

* Whether the triad is classic or mirrored depends on three factors, see the text above.

## 4. Summary

In the class of tectonic earthquakes, we found a subclass of the so-called mirror triads. A specific property of mirror triads is that, in contrast to classical triads, in which the number of aftershocks is greater than the number of foreshocks, in mirror triads the number of aftershocks is less than the number of foreshocks in the interval 24 h before and 24 h after the main shock. In many cases, there are no aftershocks at all. In addition to this, strong solitary earthquakes were discovered, which are not preceded by foreshocks, and after which there are no aftershocks.

*Acknowledgments*. We express our deep gratitude to B.I. Klain and A.D. Zavyalov for numerous discussions of problems in the physics of earthquakes. We are deeply grateful to A.S. Potapov for interest in this work. The authors thank the staff U.S. Geological Survey for providing the catalogues of earthquakes. The work was carried out according to the plan of state assignments of IPhE RAS.